\documentclass[11pt,twoside]{article}

\usepackage{asp2006}
\usepackage{epsf}
\usepackage{lscape}
\usepackage{graphicx}

\begin{document}
\title{[WC] and PG1159 Central Stars of Planetary Nebula: \\the Need for an Alternative to the Born-Again Scenario}   %%% Fill in title
\author{Orsola De Marco}   %%% Fill in author names
\affil{American Museum of Natural History}    %%% Fill in author affiliations

\begin{abstract} 
Hydrogen-deficient central stars of planetary nebula such as Wolf-Rayet and PG1159 central stars and some weak emission line stars are primarily composed of helium and carbon. This abundance is well explained by a scenario where a single post-AGB star experiences a last helium shell flash which ingests and burns, or simply dilutes, the remaining hydrogen atmosphere. But despite its success in matching the photospheric abundances of these stars, this scenario is faced with several observational challenges. A binary scenario is proposed here as a more natural way to face some of the most stringent observational constraints. In this scenario the H-rich primary in a close binary formed during a common envelope on the AGB, suffers a last helium shell flash, which results in a H-deficient primary with some of the characteristics needed to match the observations. 
\end{abstract}

%%% MAIN BODY OF TEXT GOES HERE. CONSULT "INSTRUCTIONS FOR AUTHORS USING
%%% LATEX2E MARKUP", SECTIONS 2.3-2.6 FOR HELP WITH EQUATIONS, FIGURES,
%%% AND TABLES.

\section{Introduction}  

One of many classes of H-deficient stars are the H-deficient, C-rich central stars of planetary nebula (PN). These are post-asymptotic giant branch (AGB) stars whose atmospheres are composed of helium and carbon with roughly equal mass fractions, with up to $\sim$10\% of oxygen and lower amounts of a few other metals (Werner \& Herwig 2006). Here I am not discussing other classes of H-deficient post-AGB stars (with or without PN), which are composed mostly of helium (e.g., the R Coronae Borealis stars; see Clayton, these proceedings). 

The H-deficient, C-rich central stars of PN are subdivided into a number of  sub-classes, defined by how their spectra appear. The first is the Wolf-Rayet central stars of PN (the [WC] stars\footnote{The bracket was introduced by van~der~Hucht et al. (1981), to distinguish them from massive Wolf-Rayet stars.}), whose spectral appearance is very close to that of massive Wolf-Rayet stars of the carbon sequence. These stars have broad emission lines of HeII, CII-IV, OII-OVI (Crowther et al. 1998) and are divided into the cooler, late [WCL] stars (subclasses [WC12-7]) and the hotter, earlier [WCE] stars (subclasses [WC6-4] and [WO]\footnote{The [WO] class is a natural extension of the [WCE] class to higher wind ionizations and does not reflect a change in abundances.}). Since all post-AGB stars become hotter as they evolve, it is presumed that the [WCL] stars evolve into [WCE] stars. The second class is the PG1159 stars. They have similar abundances to the [WC] stars, but have much weaker winds and hence also display absorption lines (because their weaker winds allow one to see down to the photosphere). On the Hertzsprung-Russell (HR) diagram, or the $\log g$ vs. $T_{eff}$ diagram, they occupy a location at higher gravity and lower luminosity.  Their characteristics make it plausible that the PG1159 stars are the progeny of the [WCE] stars. Only about half of the PG1159 stars have a PN (Jeffery et al. 1996, Jeffery, these proceedings), likely because, by the time a central star is a PG1159 its PN has started to disperse.

Another class of H-deficient central stars of PN, some of which are related to the [WC] and PG1159 stars (but some are H-normal and have emission lines due to stronger-than-normal winds; Mendez et al. 1990) is the weak emission line stars or WELS. The H-deficient members of this class have spectra with some week emission lines but also display photospheric absorption lines.  Fogel et al. (2003) showed that none of the analyzed PNe around WELS is of ``Type~I" (those with N/O$>$0.8, by number, which implies that they come from more massive progenitors with $M_i \ga 4$~M$_\odot$; Kingsburgh \& Barlow 1994). This could imply that these  central stars underwent a similar evolution to those of stars with stronger winds and stronger emission lines, but that they are less luminous because they are systematically less massive. As a result, their winds and emission lines are weaker. 

A few additional objects fall under the definition of H-poor central stars of PN, not necessarily because of the appearance of their stellar spectrum, but because they were witnessed to suffer an outburst with a particular set of characteristics. These are Sakurai's object (whose outburst started in 1995 and whose cool H-deficeint giant is now totally obscured by dust; Duerbeck et al. 1997), V605~Aql (the central star of the PN A~58, whose outburst started in 1917 and whose spectral type is now [WCE]; Clayton \& De~Marco 1997) and FG~Sge (which begun to brighten in 1894 and that is today a cool H-poor supergiant; Jeffery \& Schoenberner 2006). All three suffered outbursts thought to have been due to a helium shell flash on a post-AGB star. All three were observed to re-expand and re-join the AGB. One of these stars, V~605~Aql, has left the AGB once again and is now a hot central star (Clayton et al. 2006)\nocite{Clayton2006}. 

Finally, two PG1159 stars and two [WCE] star (A~30, A~78 and V605~Aql and IRAS~15154-5258; Jacoby \& Ford 1981, Manchado et al. 1989, Clayton \& De Marco 1997)\nocite{Jacoby1981,Manchado1996} also have, in addition to H-deficient central stars, prominent H-deficient ejecta inside larger, H-normal PNe, that are thought to be due to a final helium shell flash that happened a few thousands years ago.

The evolutionary scenario currently thought to explain the above stellar classes will be better explained in the next Section. With this contribution I outline the reasons why this scenario, successful as it is in explaining the abundances of these stars, has fatal flows when confronted with some other key observations. Unfortunately, the evidence against the current scenario, while being quite definitive,  is also complex and fails to present us with a clear alternative picture. This notwithstanding, there are binary scenarios that can explain at least some of the observations.

\section{The Final Thermal Pulse scenario and the formation of \\ H-deficient post-AGB stars}

At the top of the AGB, a star experiences a series of helium shell thermal pulses. These are due to an increase of the helium shell mass as a result of nuclear fusion in the hydrogen shell above.
\citet{Schoenberner1979} noticed that the inter-pulse time-scale could be shorter than the time it takes the post-AGB star to exhaust its nuclear fuel. When this is the case, the {\it last} thermal pulse happens during the pot-AGB phase. Depending on when during the post-AGB evolution this happens, the star returns to the AGB as H-poor or H-free. Eventually, the ``born-again" AGB star leaves the AGB for the second time and retraces its evolution as a H-poor/free post-AGB star. Because of the high opacity of H-depleated envelopes, the energy produced in the burning shells propels powerful winds, which exhibit the emission line spectrum of [WC] stars. 
 
\citet{Iben1983} realized that this scenario might also explain the presence of the H-deficient ejecta that had been discovered inside the two old PNe, A~30 and A78 \citep{Jacoby1981}. The abundances predicted by the born-again scenario match well those of the stellar atmospheres of [WC] and PG1159 stars (Werner \& Herwig 2006). Finally, the outburst behavior of the three born-again objects, Sakurai's object, V605~Aql and FG~Sge fits reasonably well this scenario, giving it additional support.

The evolutionary sequence suggested by this scenario is therefore the following. 
A final helium shell thermal pulse causes the star to expand again while gradually (but swiftly) losing hydrogen, the star goes through a dust-forming phase (Sakurai's object is currently in this phase) similar to that exhibited by R~Cor~Bor (RCB) stars (although, as explained earlier, RCB stars are now thought to be produced by mergers; See Clayton's contribution in these proceedings). Eventually the star heats up again and develops a [WC] spectrum, cooler at first ([WCL]), hotter later ([WCE]; this is the current spectral type of V~605~Aql). Eventually nuclear fusion stops and the star's luminosity decreases along with a gravity increase. The star is now a PG1159 star with a PN around it. Eventually the PN fades and the PG1159 star is left naked.
 
Support for this scenario, aside from the photospheric abundances, comes from the fact that the nebular properties of PNe around [WC] and PG1159 stars are by and large very similar to those of H-normal PNe. This is in line with the born-again scenario because a final thermal pulse should happen randomly to about 20\% of all post-AGB stars \citep{Gorny2000}. This statement is not absolutely true, however, and some exceptions, described below, actually contribute to the body of evidence against this scenario.   
 
\section{Observations that contradict the born-again scenario for the formation of all H-deficient post-AGB stars}

%In this Section I discuss some observational evidence against a born-again origin for all H-deficient post-AGB stars. This does not mean that a final thermal pulse cannot happen. 
Final thermal pulses {\it must} happen to a subset of post-AGB stars. But it is clear that, in its current formulation, this scenario cannot be the {\it only} explanation for H-deficient central stars of PN.

\subsection{The ejecta in the born again PNe A~30 and A~58}

In the born-again scenario, H-deficient knots are ejected during the final thermal pulse event and are therefore predicted to have abundances similar to those of the H-deficient photosphere of the born-again star.
The H-deficient knots of two born-again central stars of PN, A~30 and A~58, have been analyzed using optical recombination lines (ORLs) instead of collisionally-excited lines (CELs). These analyses
have revealed that the H-deficient ejecta are in both stars oxygen-rich (mass fractions are: C=6\%, O=24\% for A~30's knots (average value) and 2\% and 32\% for A~58's knot; \nocite{Wesson2003}, Wesson et al. (2003,2007)). In addition, the same analyses revealed a very large abundance of neon (34\% and 13\% for A~30 and A~58's knots, respectively). These abundance patterns are at odds with those predicted by the born again scenario (C$\sim$40-30\%, He$\sim$60-30\%, O$\sim$few-20\% and a neon mass fraction $\la$2\%; \citealt{Werner2006}). The abundances are only as uncertain as the atomic data which are used to determined them. This uncertainty, even if taken  conservatively, cannot reconcile the abundances with the predictions, and provide a formidable challenge to the born-again scenario for the evolution of these two key objects. These abundances were compared by Wesson et al. (2007) to those of nova ejecta, in particular the ejecta of oxygen-neon-magnesium novae. It is not straight forward to suggest that these two born-again stars are ONeMg novae, but it is clear that a critical evaluation of these abundances is needed, because they are in glaring contrast with the predictions of the final thermal pulse scenario.

It is interesting to notice that the H-deficient knots more readily shine in the ORLs because this radiation is favoured by the cool dense conditions prevailing in the knots. Many PNe around H-rich {\it and} H-deficient central stars, have very different abundances when probed by ORL or CEL radiation \citep{Liu2000}. This discrepancy can be explained if these PNe contain small H-deficient knots (in most cases too small to be imaged directly). If this is the case, H-deficient ejecta would be a more common and complex phenomenon that can be accounted for by our current formulation of the born again scenario. 

%One might wonder whether these analyses might have incorrectly assessed the PN and ejecta abundances. This would be a fair question in that the abundances of PNe have recently been put in doubt. It has been revealed that the abundances of ionized low density gas determined from CELs (the so called forbidden lines) can be systematically different from those determined from ORLs \citep[e.g.,]{Wesson2005}. The latter are much fainter and they can be measured only in the brightest PNe. The source of the discrepancy, it is thought, lies in the fact that there are two components in the plasma: one is the lower density, hydrogen-rich PN, which predominately shines in the forbidden lines and hence can be probed by them. The second is a denser, cooler hydrogen-poor component in the form of small knots, which shine primarily in recombination lines because of their high density. 

%The oxygen, carbon and neon abundances on the hydrogen-poor knots of A~30 and A~58 would be those derived from the ORLs. Wesson et al. (2003,2007) showed that indeed these knots have very week CEL light and that it is the ORLs that provide a probe of their conditions. If this is so, then the abundances are only as uncertain as the atomic data which are used to determined them. This uncertainty, even if taken  conservatively, cannot reconcile the abundances with the predictions, and provide a formidable challenge to the born-again scenario for the evolution of these two objects.

The suggestion of a ONeMg nova scenario for these two born-again stars, is entirely based on the abundance of the ejecta as there is at present no other evidence to substantiate the connection. However, there are a few seemingly disconnected pieces of evidence, which might one day prove fundamental in linking born-again stars and novae. (i) The hydrogen-deficient ejecta of A~30 and A~78 have collimated polar knots, hard to achieve unless an accretion disk exists or has existed around one of the components of a close binary\footnote{The fact that the older, H-rich PNe around these two objects are circular, or at most elliptical, has often been used as evidence that the progenitor stars are single. However, it must be bourn in mind that morphology is not yet understood enough to be used as a certain proof of stellar origin (the debate over what shapes PNe has raged for over 30 years; \citealt[][and references therein]{DeMarco2006}).}.
(ii) The Nova CK~Vul is today thought by some to have suffered a born-again outburst in view of its H-deficient ejecta. Others, however, believe that this is a more or less regular nova, {\it despite} its H-deficeint ejecta \citep[][and references therein]{Hajduk2007}. There is no optical central star in the middle of this nebula. Although the jury is still out, this object is likely to provide a link between novae and born-again stars. (iii) A PG1159 star, though one without a visible PN, has recently been found to be in a close binary connecting binarity and H-deficient PNe for the first time \citep{Nagel2006}.

\subsection{The [WCL] stars: a bag of oddities}

\subsubsection*{The dual-dust chemistry.}

All the well-observed [WCL] central stars of PN exhibit carbon- (polycyclic aromatic hydrocarbons: PAHs) and oxygen rich (crystalline silicates) dust {\it near} the central star (see \citet{DeMarco2002} for the statistics). Almost no [WCE], the suggested descendants of the [WCL] stars, has the same characteristic (one object only: NGC~5315; \citealt{DeMarco2002}). PG1159 stars, the supposed next link in the evolutionary chain, are not known to exhibit the dual-dust chemistry, although they have not been systematically checked. There is no other known central star of PN which exhibit dual-dust chemistry circumstellar environments\footnote{H-normal post-AGB stars (with no PN) in binaries do systematically exhibit the dual dust chemistry. \citet{vanWinckel2001} suggests that these are the progenitors of the [WC] central stars. While I recognize that there is probably a connection, we cannot simply state they are the progenitors, since we would have to explain why all of these post-AGB binaries, but only them, go through a born-again evolution.}.

%The only other objects that show the simultaneous presence of oxygen and carbon rich dusts are some massive, oxygen rich stars, the ejecta of some novae and some post-AGB stars (these stars have left the AGB but are not yet hot enough to ionized the surrounding mass-loss and display their PN). The first two cases are thought to have oxygen-rich (gaseous) outflows, where silicate dusts can form. The strong UV fields of the central sources then dissociate enough CO molecules to make some carbon available for the formation of carbon-rich dusts (PAHs). These objects are unlikely to be related to the dual-dust [WC] stars. The third category is instead thought to have massive dusty, oxigen-rich disks that formed during the early AGB by the action of a companion at some intermediate distance. When the AGB star started to undergo thermal pulses its mass-loss outflow became C-rich and started making PAHs. The companions are today still detectable at close-to-intermediate separations (100$<$P$<$1500~days; \citealt{vanWinckel2003}).  
\begin{figure}[!ht]
%\vspace{2cm}
\includegraphics[bb = -250 20 33 400, scale=0.3]{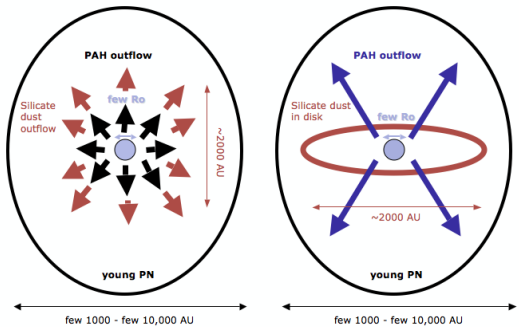}
%\plotone{DualDustCartoon.eps}
\caption{A cartoon representing current explanations of the dual-dust chemistry observation in [WCL] central stars of PN. Left: the recent AGB departure scenario envisages that the O-rich outflow was followed by the C-rich outflow, by AGB departure {\it and} by a final thermal pulse, in quick succession. Right: the disk storage scenario relaxes the timing constraint by envisaging tha§ t the O-dust is stored in an orbiting disk.}
\label{fig:DualDustCartoon}
\end{figure}

The presence of near-star warm oxygen- and carbon-rich dust has been explained  by  suggesting that the [WC] star was an O-rich AGB star only a few thousands years ago \citep[][Fig.~\ref{fig:DualDustCartoon}, left]{Waters1998}. Only so would the O-rich dust that condensed in the flow be still near the star. Within this period of time the chemistry of the outflow changed to carbon-rich due to the thermal pulses {\it and} the star departed from the AGB. Even if the change in chemistry could have triggered a swift departure from the AGB, it is unclear why all these objects then became [WCL] stars and why there aren't some H-rich central stars also associated with the dual-dust. %In other words, the fact that the dual-dust is {\it exclusively} associated with [WC] stars likely indicates that they have a common cause. %whether such fine tuning could happen to enough stars. The presence of a [WC] stars also of course imposes the requirement that these finely-tuned objects must have undergone a final thermal pulse. This would mean that not only the change in chemistry dictated AGB departure, but also that it caused a thermal pulse. 

\begin{figure}[!ht]
\includegraphics[bb = -320 30 33 270, scale=0.4]{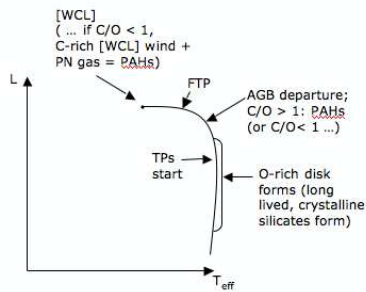}
\caption{A cartoon of the HR diagram graphically explaining the disk storage scenario for the existence of dual-dust chemistry [WCL] stars. The silicates are formed at the time when the early AGB star is O-rich and are stored in a large disk. The star can later become C-rich as the result of the third dredge up which follows thermal pulses. This C-rich outflow can form PAHs. The star later suffers a final thermal pulse and becomes a [WC] central star.}
\label{fig:DiskStorageScenario}
\end{figure}

\citet{Cohen1999}  later proposed that the oxygen-rich dust is stored in a large orbiting torus (Fig.~\ref{fig:DualDustCartoon}, right, and Fig.~\ref{fig:DiskStorageScenario}). This relaxes the constraint that the AGB progenitor of the [WC] stars must have been O-rich in the recent past. Yet, this scenario, like its predecessor,  does not explain why a final thermal pulse, which makes the star H-deficient, should happen {\it only} to those objects that have formed an O-rich disk during the early AGB.
Both these scenarios should form hydrogen-normal as well as hydrogen-poor dual-dust central stars of PN. In other words, the dual-dust chemistry property should be, as is the case for many other properties of the PNe around [WC] central stars, similarly seen in H-rich and H-poor objects. 

Finally both in the \citet{Waters1998} and the \citet{Cohen1999} scenarios the reason why [WCE] and PG1159 stars do not have dual-dust chemistry is that the rising temperature of these stars destroys the dust. It is not clear that the crystalline silicates present in the oxygen rich disks can be obliterated easily, and probably not in a matter of a few thousands years, which is the time that separates the [WCL] from the [WCE]/PG1159 stars.

\subsubsection*{Frequency of the [WC] central stars in different environments.}

The H-deficient central star phenomenon, is clearly more complex than one that it brought about by a final thermal pulse. Another example of this is that in different environments we find different [WC] central star fractions. There are 4 known PNe in the Sagittarius Dwarf Galaxy. Three are [WC] central stars \citep{Zijlstra2006} (cfr. with about 10\% in the Galaxy)! This is possibly an indication that metallicity has something to do with the formation of a [WC] star. 

Second, \citet{Gorny2004} found a larger fraction of [WCL] central stars in the bulge than it is known in the Galaxy in general. This might indicate a metallicity dependence of the [WC] subclass distribution. This is not unlikely in view of the fact that two H-deficient post-AGB stars, with the same core mass and at the same evolutionary stage, would show different [WC] wind properties if they had different metallicities (see Crowther, this proceedings).  

Finally, of the four PNe known to reside in globular clusters, one is H-deficient.  This object is very different from the usual H-deficient PN, such as A~30, A~78 or even IRAS15154-5258, primarily because the central star is H-rich \citep{Harrington1996}. This object is likely related to a merger event since its high luminosity indicates that the central star mass is too large to be coeval with the old population of its home globular cluster (M~22). This object establishes a connection between mergers and H-deficient ejecta and shows that H-deficiency might not be the prerogative of the final thermal pulse scenario.

\subsubsection*{Frequency of the [WCL] central stars compared to [WCE] and PG1159 stars.}
\citet{Zijlstra2001} discusses properties of [WCL] central stars that, once again, set them aside from other H-deficient central stars. 
%In particular their IRAS fluxes and colors position the [WCL] stars between post-AGB stars and hotter central stars of PN. This location of the IRAS color-color and color-magnitude diagram, seems logical for these young central stars, but is actually a particularly underpopulated area of these diagrams. H-rich central stars of PN, when they are young, have significant amounts of hot dust and indeed appear close to the post-AGB stars in the IRAS color-color diagram. However, this condition is very short lived because the star quickly evolves to higher temperature and destroys the dust. As a result there are relatively very few H-rich central stars with bright and blue IRAS fluxes. On the other hand, \citet{Zijlstra2001} argues, t
There are too many [WCL] central stars relative to the [WCE] and PG1159 stars,
% occupying this region of the IRAS color-color diagram. Indeed there are too many [WCL] objects in general, 
in particular  when we consider that these stars must have a very low envelope mass compared to H-rich central stars {\it and} they have mass-loss rates which are much higher. Since leftward evolution on the HR diagram depends on envelope mass, the [WCL] central stars  should be moving to the hot side of the HR diagram with breakneck speed, much faster than their H-rich counterparts. \citet{Zijlstra2001} suggests that the [WCL] are stalled in their leftward motion on the HR diagram by accretion of circumstellar material.

\subsubsection*{A comparison between the PNe around [WC] star dynamical ages and their central stars evolutionary speeds.}

\citet{Zijlstra2006} carried out an experiment where he measured central star masses by determining their effective temperature and their PN dynamical age. The assumption is that the PN dynamical age is the same as the time since the star left the AGB and hence a measure of how quickly it reached its current position on the HR diagram as determined by its temperature. This speed is also a measure of the stellar mass since evolutionary speed has a very strong dependence on stellar mass. This method, they claim, is quite accurate since even if the determination of the dynamical age of a PN is inaccurate, the steep mass-evolutionary speed relationship insures a relatively low error on the central star mass.

Applying this method they determined a central star of PN mass distribution peaking at the same mass as pervious determinations ($\sim$0.62~M$_\odot$), but with a much smaller standard deviation. They also determined that the mass distribution of the [WCE] central stars peaks at the same mass but it is narrower still, while the [WCL] central stars have a mass distribution that is totally flat, occupying both the lowest and highest mass bins.  

This is hard to explain. It is unlikely that what forms [WC] stars is restricted to a specific mass bin. 
%It would make some sense {\it not} to have particularly massive [WC] stars because those would evolve too fast for a final thermal pulse to actually take place before nuclear fusion stops. But it makes no sense that these stars all cluster at some intermediate mass. 
In fact, one would actually predict the distribution of [WC] masses determined with the method above to tend towards lower masses: the final thermal pulse will carry the central star back to the AGB and then once more to the hot part of the HR diagram, all the while the PN would be expanding. This means that the PN should always be too large for the position of the [WC] star on the HR diagram, so that one would conclude that the central star was a slow evolver, and hence has low mass. This is not the case for the [WCLs] which span the entire range and is not the case for the [WCE] stars which cluster in the middle of the mass range.

%In conclusion, while the spread in masses of the [WCL] might be explainable (see below) the concentration of the [WCE] on one mass bin is extremely puzzeling. If this result holds, it will be an interesting one to be explained by any theory of [WC] star evolution.

\section{Binary scenarios to explain H-deficient post-AGB stars}

The scenarios that follow do not pretend to address all of the observations listed above. They primarily address the dual-dust chemistry phenomenon. In particular, they address the fact that unless  H-deficiency is caused {\it by} the dual-dust or the dual-dust {\it and} the H-deficiency have a {\it common} cause, there is no way to explain the absence of H-rich dual-dust chemistry central stars of PN. In what follows we therefore seek a {\it causal link} either between the dual-dust and the H-deficiency, or, alternatively, we look to find a common cause of both. This common cause is assumed in these scenarios to be the penetration of the AGB or post-AGB star by a companion.

\subsection*{The merger scenario.}
\begin{figure}[!ht]
\includegraphics[bb = -120 30 33 270, scale=0.4]{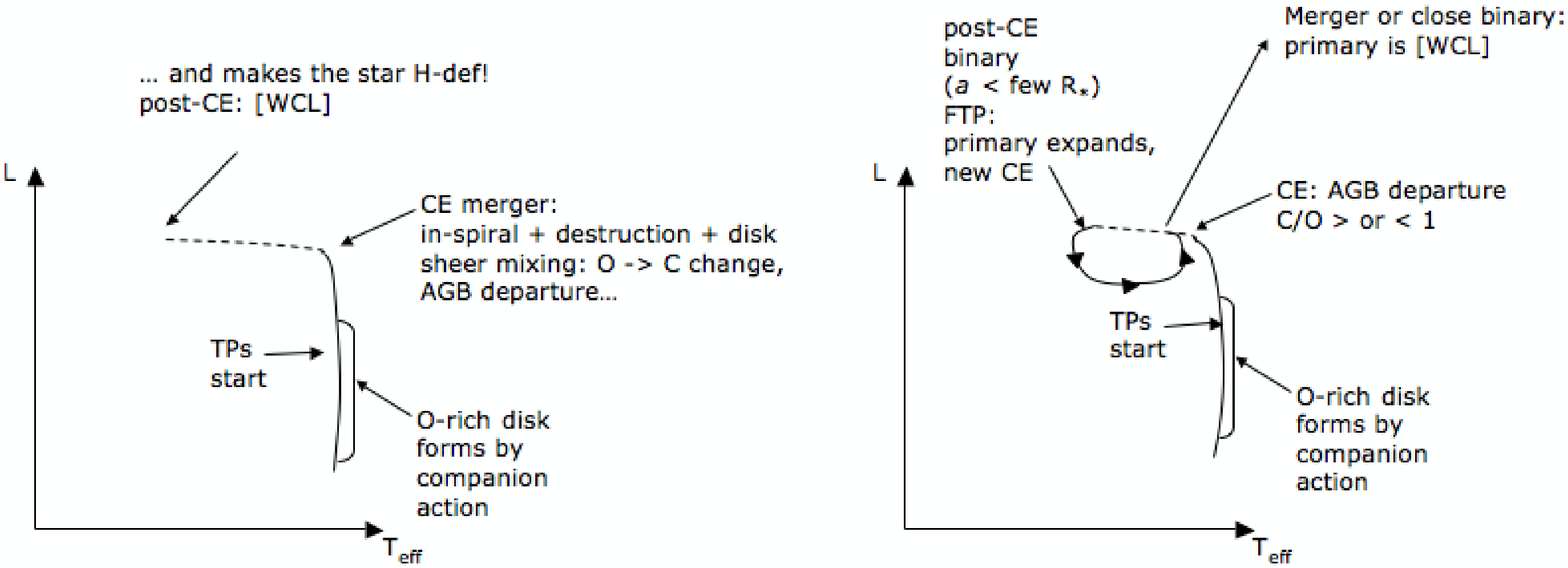}
\caption{A cartoon HR diagram graphically explaining two binary scenarios for the formation of dual-dust chemistry [WCL] stars. Left: the merger scenario of De~Marco \& Soker (2002), where a small companion merges with the AGB star causing the chemistry change, the departure from the AGB and the H-deficiency. Right: here a common envelope on the AGB forms a close binary with an H-rich central star. Later a final thermal pulse turns the primary into a [WCL] and interaction with the companion or its shredded remains, facilitates PAH formation.}
\label{fig:BinaryScenarios}
\end{figure}

De~Marco \& Soker (2002) considered the dual-dust properties of the [WCL] stars and suggested a binary scenario for the evolution of these objects. In this scenario (Fig.~\ref{fig:BinaryScenarios}, left panel) a small companion enters common envelope with the primary on the AGB. Because if its small mass it continues to spiral into the primary's envelope till, at about 0.1~R$_\odot$ it gets tidally destroyed and forms a disk just above the two burning shells. This disk dredges up C-rich material from the inter-shell region by sheer mixing and changes the chemistry from oxygen to carbon rich. It also stimulates a mass-loss enhancement, maybe via enhancing dust formation and causes departure from the AGB. If enough mass-loss can be caused by this intrusion, the resulting post-AGB star, and eventually the central star of the ensuing PN, might be H-deficient. This scenario's main flow is likely to reside in the fact that it is extremely hard for an AGB star, even one at the top of the AGB with a relatively small envelope mass, to just lose all the remaining hydrogen. This notwithstanding, outlining this scenario serves, aside from giving the historical roots to the binary idea for these stars, to exemplify what I mean by a cause of both the dual-dust chemistry {\it and} the H-deficiency. What follows might be a more viable scenario.

\subsection*{The two-common envelope scenario.}

%A variation of this scenario would be to add the requirement that in order for the common envelope to dictate AGB departure the companion needs to emerge from the common envelope. Suggesting the presence of a companion that enters common envelope at some point in the upper AGB, can also naturally produce an O-rich disk during the early AGB. A close binary emerges from the common envelope with the primary at some hotter temperature (Fig.~\ref{fig:binaryscenarios}, right). The primary is at this point not H-deficient. However if the post-common envelope binary were close enough, the companion could fill its Roche Lobe and start mass transfer onto the primary right away. If the transfer rate were high enough the primary would expand and enter a second common envelope which, this time, could result in a merger. Due to the low envelope of the primary the merger could induce H-deficiency, or alternatively, could stimulate a final thermal pulse which would result in H-deficiency.  

One must wonder what happens to those post-common envelope central star binaries which undergo a final thermal pulse. Today we know of about a dozen close binary central stars with periods smaller than 3 days, H-normal primaries and cool, main sequence companions \citep{DeMarco2006}. As is the case for single stars, about 20\% of the primaries in these systems should suffer a final thermal pulse. It is not obvious what would happen to these systems when the primary expands under the influence of the outburst. Does the companion, already so close to the primary, merge, or does the companion  survive? %It is also not entirely clear whether, in the case of a merger, the resulting single star would be H-poor, since the companion is made of almost pure hydrogen and has a mass between 0.2 and 0.5~M$_\odot$ which would recoat the central star with hydrogen.

Here we speculate that these systems originate the dual-dust chemistry [WCL] central stars. The oxygen rich dust would be in a disk as in the disk-storage scenario. In the case of binaries, which eventually enter common envelope in the upper AGB, the formation of a disk during the lower, O-rich, AGB phase is  plausible. As for the C-rich dust, the condition that this binary scenario (and indeed {\it any} scenario) must satisfy, is that the cause for the dual-dust chemistry is the very binary interaction that causes H-deficiency. If this is not so, a binary scenario would have the usual problem: it should generate dual-dust chemistry also around some H-rich central stars. 

It is possible that once the final thermal pulse happens in these close binary systems, the presence a companion so close to the H-deficient born-again central star, facilitates PAH formation. In the case of a merger, perhaps, the companion material provides the H-rich gas that, together with C-rich wind of the newly-forged [WCL] star, permits the formation of PAHs. Maybe the newly formed H-deficient central star will accrete material from its surrounding H-rich material and be temporarily stalled as was suggested by Zijlstra (2001). And maybe the hydrogen rich material will indeed re-coat the central star which would then become H-rich once again (as is the case for IRAS18333-2357 wich is H-rich in the middle of H-poor ejecta; Harrington 1996). 

In this scenario there is room for single stars which undergo a final thermal pulse. These tend to have PAHs which condense in the C-rich AGB flow or, possibly, during the post-AGB [WC] phase as the C-rich winds impact H-rich nebular material. In these single stars there would be no dual-dust chemistry because the disk would not have formed. These single stars would rarely, if at all, be seen as [WCL] because this phase is very fast. They would all become [WCE] and PG1159 stars. From the frequency of post-common envelope central stars ($\sim$10-15\%; \citet{Bond2000}) and the fact that about 20\% of them should suffer a final thermal pulse, we can predict that 2-3\% of all central stars should be dual-dust [WCL] stars. While this very approximate figure is in line with the known fraction of [WCL] central stars, we should point out that the central star binary fraction of 10-15\% is a lower limit restricted to the very shortest period binaries. Also, as the number of free parameters in a binary is large, we have no idea which subset of these systems actually triggers the dual-dust chemistry. For instance, one could also envisage a way in which some post-outburst central stars binaries with slightly wider companions, could return to the AGB and later become the post-AGB, H-rich, binaries with dual dust discovered by van~Winckel (2001).

No matter, I think that this suggestion is intriguing because we know that a final thermal pulse must happen in post-common envelope central stars and these already have the pre-requisites to have formed an O-rich disk. If a reasonable mechanism is found by which these systems make strong PAHs, then the condition of the exclusive connection between dual-dust and H-deficiency would be satisfied.

\section{Conclusions}
\label{sec:conclusions}
 
It is not lightly that one should introduce complexity into a scenario that is quite elegant and simple. However there is reason to believe that not all H-deficeint central stars of PN derive from a born-again event which makes them into [WCL], then [WCE] and finally PG1159 stars. We propose instead that born-again events do happen and when they do they generate [WC] stars which however evolve too rapidly towards the hot regime in the HR diagram, to ever be observed as [WCL] stars. They are instead observed as hot [WCE] or, directly, as PG1159 stars. The [WCL] stars (or in any case the bulk of them) are produced by another scenario, maybe one that involves a binary interaction and which plays a role in the formation of the dual-dust. These objects remain [WCL] stars for a longer time, because of re-accretion of near-star common envelope material. They eventually move fast to the hot part of the HR diagram becoming [WCE] for a short time (there is one known [WCE] with dual-dust which could descend from the [WCL] stars) and eventually PG1159 stars. If this is so, then we predict that some PG1159 stars should have dual-dust. Alternatively, the H-rich material accreted re-coats the star which then becomes H-normal again (as could be the case for IRAS18333-2357 in M22) and possibly the case for all those H-rich central stars with H-deficient ejecta inferred from the ORL abundances.

This split scenario does not explain how a mass-losing [WCL] star can possibly be accreting matter, although this is not an impossibility \citep{Soker2002}. It also does not entirely explain the evolution of the H-deficeint central stars with neon-rich ejecta, although the variety of parameters could result in several variations on a theme for the production of H-deficient central stars.

Finally, these stars should be set in the wider context of PNe. After a three-decade debate, it has finally been accepted in the light of new theoretical evidence that indeed a binary is likely to be needed to shape non-spherical PNe (Soker 2006, Nordhaus et al 2007).  While this theoretical claim has not yet been substantiated by observations (for a review see De~Marco 2006), one should keep in mind the possibility that binaries might interfere with single star evolution in the majority of cases. If this is so, the number of possible evolutionary scenarios is much larger, since the possible combinations of a final thermal pulse with the binary stellar and system parameters are many, many more than the parameters of a single stars. In such a setting, the number of possible futures of any AGB star is raised exponentially.   

\acknowledgements  
It is a pleasure to acknowledge many conversations with my friends and colleagues Falk Herwig, Noam Soker, Howard Bond, George Jacoby and Geoff Clayton and other members of the Planetary Nebula Binaries (PLAN-B) working group (www.wiyn.org/planb/). This research was partly supported by National Science Foundation Grant AST-0607111. 

%%% THE BIBLIOGRAPHY
%%%
%%% CONSULT SECTION 3 OF "INSTRUCTIONS FOR AUTHORS" FOR HOW TO USE NATBIB.
%%% AUTHORS ARE ENCOURAGED TO USE EITHER THE "THEBIBLIOGRAPY" ENVIRONMENT
%%% BY UNCOMMENTING (DELETING THE "%" SYMBOL) THE COMMANDS BELOW, OR BY
%%% USING THE BIBTEX ENVIRONMENT. TO FIND OUT WHICH IS APPLICABLE TO YOUR
%%% CONTRIBUTION, CONSULT THE VOLUME EDITORS FOR YOUR PROCEEDINGS.
%%%
\nocite{DeMarco2006}
\nocite{Werner2006}
\nocite{vanderHucht1981}
\nocite{Crowther1998}
 \nocite{Mendez1990}
 \nocite{Fogel2003}
 \nocite{Kingsburgh1994}
 \nocite{Clayton1997}
 \nocite{Duerbeck1997}
 \nocite{Jeffery2006}
  \nocite{Jeffery1996}
 \nocite{Manchado1989}

%\bibliographystyle{/Users/orsola/Work/apj}                       %% AASTeX
%\bibliography{/Users/orsola/Work/bibliography}

\begin{thebibliography}{}
%\expandafter\ifx\csname natexlab\endcsname\relax\def\natexlab#1{#1}\fi

\bibitem[{{Bond}(2000)}]{Bond2000}
{Bond}, H.~E. 2000, in ASP Conference Series, Vol. 199,
  Asymmetrical Planetary Nebulae II: From Origins to Microstructures, ed. J.H.~{Kastner},
  N.~{Soker}, \& S.~{Rappaport}, 115--+


\bibitem[{{Clayton} \& {De Marco}(1997)}]{Clayton1997}
{Clayton}, G.~C. \& {De Marco}, O. 1997, \aj, 114, 2679

\bibitem[{{Clayton} {et~al.}(2006){Clayton}, {Kerber}, {Pirzkal}, {De Marco},
  {Crowther}, \& {Fedrow}}]{Clayton2006}
{Clayton}, G.~C., {Kerber}, F., {Pirzkal}, N., {De Marco}, O., {Crowther},
  P.~A., \& {Fedrow}, J.~M. 2006, \apjl, 646, L69

\bibitem[{{Cohen} {et~al.}(1999){Cohen}, {Barlow}, {Sylvester}, {Liu}, {Cox},
  {Lim}, {Schmitt}, \& {Speck}}]{Cohen1999}
{Cohen}, M., {Barlow}, M.~J., {Sylvester}, R.~J., {Liu}, X.-W., {Cox}, P.,
  {Lim}, T., {Schmitt}, B., \& {Speck}, A.~K. 1999, \apjl, 513, L135

\bibitem[{{Crowther} {et~al.}(1998){Crowther}, {De Marco}, \&
  {Barlow}}]{Crowther1998}
{Crowther}, P.~A., {De Marco}, O., \& {Barlow}, M.~J. 1998, \mnras, 296, 367

\bibitem[{{De Marco}(2006)}]{DeMarco2006}
{De Marco}, O. 2006, ArXiv Astrophysics e-prints

\bibitem[{{De Marco} \& {Crowther}(1998)}]{DeMarco1998}
{De Marco}, O. \& {Crowther}, P.~A. 1998, \mnras, 296, 419

\bibitem[{{De Marco} \& {Soker}(2002)}]{DeMarco2002}
{De Marco}, O. \& {Soker}, N. 2002, \pasp, 114, 602

\bibitem[{{Duerbeck} {et~al.}(1997){Duerbeck}, {Benetti}, {Gautschy}, {van
  Genderen}, {Kemper}, {Liller}, \& {Thomas}}]{Duerbeck1997}
{Duerbeck}, H.~W., {Benetti}, S., {Gautschy}, A., {van Genderen}, A.~M.,
  {Kemper}, C., {Liller}, W., \& {Thomas}, T. 1997, \aj, 114, 1657

\bibitem[{{Fogel} {et~al.}(2003){Fogel}, {De Marco}, \& {Jacoby}}]{Fogel2003}
{Fogel}, J., {De Marco}, O., \& {Jacoby}, G. 2003, in IAU Symposium, Vol. 209,
  Planetary Nebulae: Their Evolution and Role in the Universe, ed. S.~{Kwok},
  M.~{Dopita}, \& R.~{Sutherland}, 235--+

\bibitem[{{G{\'o}rny} {et~al.}(2004){G{\'o}rny}, {Stasi{\'n}ska}, {Escudero},
  \& {Costa}}]{Gorny2004}
{G{\'o}rny}, S.~K., {Stasi{\'n}ska}, G., {Escudero}, A.~V., \& {Costa},
  R.~D.~D. 2004, \aap, 427, 231

\bibitem[{{G{\'o}rny} \& {Tylenda}(2000)}]{Gorny2000}
{G{\'o}rny}, S.~K. \& {Tylenda}, R. 2000, \aap, 362, 1008

\bibitem[{{Hajduk} {et~al.}(2007){Hajduk}, {Zijlstra}, {van Hoof}, {Lopez},
  {Drew}, {Evans}, {Eyres}, {Gesicki}, {Greimel}, {Kerber}, {Kimeswenger}, \&
  {Richer}}]{Hajduk2007}
{Hajduk}, M., {Zijlstra}, A.~A., {van Hoof}, P.~A.~M., {Lopez}, J.~A., {Drew},
  J.~E., {Evans}, A., {Eyres}, S.~P.~S., {Gesicki}, K., {Greimel}, R.,
  {Kerber}, F., {Kimeswenger}, S., \& {Richer}, M.~G. 2007, \mnras, 378, 1298

\bibitem[{{Harrington}(1996)}]{Harrington1996}
{Harrington}, J.~P. 1996, in ASP Conf. Ser. 96: Hydrogen Deficient Stars,
  193--+

\bibitem[{{Iben} {et~al.}(1983){Iben}, {Kaler}, {Truran}, \&
  {Renzini}}]{Iben1983}
{Iben}, I., {Kaler}, J.~B., {Truran}, J.~W., \& {Renzini}, A. 1983, \apj, 264,
  605

\bibitem[{{Jacoby} \& {Ford}(1981)}]{Jacoby1981}
{Jacoby}, G.~H. \& {Ford}, H.~C. 1981, in Bulletin of the American Astronomical
  Society, Vol.~13, Bulletin of the American Astronomical Society, 854--+

\bibitem[{{Jeffery} \& {Sch{\"o}nberner}(2006)}]{Jeffery2006}
{Jeffery}, C.~S. \& {Sch{\"o}nberner}, D. 2006, \aap, 459, 885

\bibitem[{{Jeffery} {et~al.}(1996){Jeffery}, {Heber}, {Hill}, {Dreizler}, {Drilling}, {Lawson}, {Leuenhagen} \&
  {Werner}}]{Jeffery1996}
{Jeffery}, C.~S., {Heber}, U., {Hill}. P.~W., {Dreizler}, S., {Drilling}, J.~S., {Lawson}, W.~A., {Leuenhagen}, U., \& {Werner}, K. 1996, in ASP Conf. Ser. 96: Hydrogen Deficient Stars,
  471--+


\bibitem[{{Kingsburgh} \& {Barlow}(1994)}]{Kingsburgh1994}
{Kingsburgh}, R.~L. \& {Barlow}, M.~J. 1994, \mnras, 271, 257

\bibitem[{{Liu} {et~al.}(2000){Liu}, {Storey}, {Barlow}, {Danziger}, {Cohen} \&
  {Bryce}}]{Liu2000}
{Liu}, X.-W., {Storey}, P.~J., {Barlow}. M.~J., {Danziger}, I.~J., {Cohen}, M. \& {Bryce}, M. 2000, \mnras, 312, 585

\bibitem[{{Manchado} {et~al.}(1989){Manchado}, {Garcia-Lario}, \&
  {Pottasch}}]{Manchado1989}
{Manchado}, A., {Garcia-Lario}, P., \& {Pottasch}, S.~R. 1989, \aap, 218, 267

\bibitem[{{Mendez} {et~al.}(1990){Mendez}, {Herrero}, \&
  {Manchado}}]{Mendez1990}
{Mendez}, R.~H., {Herrero}, A., \& {Manchado}, A. 1990, \aap, 229, 152

\bibitem[{{Nagel} {et~al.}(2006){Nagel}, {Schuh}, {Kusterer}, {Stahn},
  {H{\"u}gelmeyer}, {Dreizler}, {G{\"a}nsicke}, \& {Schreiber}}]{Nagel2006}
{Nagel}, T., {Schuh}, S., {Kusterer}, D.-J., {Stahn}, T., {H{\"u}gelmeyer},
  S.~D., {Dreizler}, S., {G{\"a}nsicke}, B.~T., \& {Schreiber}, M.~R. 2006,
  \aap, 448, L25
  
  \bibitem[{{Nordhaus}{et al.}(2007){Blackman}{Frank}}]{Nordhaus2007}
{Nordhaus}, J., {Blackman}, E.G., \& {Frank}, A. 2007, \mnras, 376, 599

\bibitem[{{Schoenberner}(1979)}]{Schoenberner1979}
{Schoenberner}, D. 1979, \aap, 79, 108

\bibitem[{{Soker}(2002)}]{Soker2002}
{Soker}, N. 2002, \mnras, 330, 481

\bibitem[{{Soker}(2006)}]{Soker2006}
{Soker}, D. 2006, \pasp, 118, 260


\bibitem[{{van der Hucht} {et~al.}(1981){van der Hucht}, {Conti}, {Lundstrom},
  \& {Stenholm}}]{vanderHucht1981}
{van der Hucht}, K.~A., {Conti}, P.~S., {Lundstrom}, I., \& {Stenholm}, B.
  1981, Space Science Reviews, 28, 227

\bibitem[{{Van Winckel}(2001)}]{vanWinckel2001}
{Van Winckel}, H. 2001, \apss, 275, 159

\bibitem[{{Waters} {et~al.}(1998){Waters}, {Beintema}, {Zijlstra}, {de Koter},
  {Molster}, {Bouwman}, {de Jong}, {Pottasch}, \& {de Graauw}}]{Waters1998}
{Waters}, L.~B.~F.~M., {Beintema}, D.~A., {Zijlstra}, A.~A., {de Koter}, A.,
  {Molster}, F.~J., {Bouwman}, J., {de Jong}, T., {Pottasch}, S.~R., \& {de
  Graauw}, T. 1998, \aap, 331, L61

\bibitem[{{Werner} \& {Herwig}(2006)}]{Werner2006}
{Werner}, K. \& {Herwig}, F. 2006, \pasp, 118, 183

\bibitem[{{Wesson} {et~al.}(2007){Wesson},{Barlow}, {Liu}, {Ercolano} \& {De Marco}}]{Wesson2007}
{Wesson}, R., {Barlow}, M.~J., {Liu}, X.-W., {Ercolano}, B., \& {De Marco}, O. 2007, \mnras, in press

\bibitem[{{Wesson} {et~al.}(2003){Wesson}, {Liu}, \& {Barlow}}]{Wesson2003}
{Wesson}, R., {Liu}, X.-W., \& {Barlow}, M.~J. 2003, \mnras, 340, 253

\bibitem[{{Zijlstra}(2001)}]{Zijlstra2001}
{Zijlstra}, A.~A. 2001, \apss, 275, 79

\bibitem[{{Zijlstra} {et~al.}(2006){Zijlstra}, {Gesicki}, {Walsh},
  {P{\'e}quignot}, {van Hoof}, \& {Minniti}}]{Zijlstra2006}
{Zijlstra}, A.~A., {Gesicki}, K., {Walsh}, J.~R., {P{\'e}quignot}, D., {van
  Hoof}, P.~A.~M., \& {Minniti}, D. 2006, \mnras, 369, 875

\end{thebibliography}

%\bibitem[]{}
%\bibitem[]{}
%\bibitem[]{}
%\bibitem[]{}
%\bibitem[]{}
%\bibitem[]{}
%\bibitem[]{}
%\bibitem[]{}
%\bibitem[]{}
%\bibitem[]{}
%\bibitem[]{}
%\bibitem[]{}
%
%\subsection{}   %%% Second level section head (remove "%" symbol)
%\subsubsection{}   %%% Lowest level section head (remove "%" symbol)
%\section*{}    %%% Unnumbered top level section head (remove "%" symbol)
%\subsection*{}   %%% Unnumbered second level section head (remove "%" symbol)

\end{document}